\documentclass[10pt, a4paper]{article}
\usepackage{lrec2022} 
\usepackage{multibib}
\newcites{languageresource}{Language Resources}
\usepackage{graphicx}
\usepackage{tabularx}
\usepackage{soul}
\usepackage{multirow}
\usepackage{titlesec}
\titleformat{\section}{\normalfont\large\bfseries\center}{\thesection.}{1em}{}
\titleformat{\subsection}{\normalfont\SmallTitleFont\bfseries\raggedright}{\thesubsection.}{1em}{}
\titleformat{\subsubsection}{\normalfont\normalsize\bfseries\raggedright}{\thesubsubsection.}{1em}{}
\renewcommand\thesection{\arabic{section}}
\renewcommand\thesubsection{\thesection.\arabic{subsection}}
\renewcommand\thesubsubsection{\thesubsection.\arabic{subsubsection}}

\usepackage{epstopdf}
\usepackage[utf8]{inputenc}

\usepackage{hyperref}
\usepackage{xstring}

\usepackage{color}

\usepackage{CJKutf8}
\newcommand{\jp}[1]{ \begin{CJK}{UTF8}{ipxm}#1\end{CJK} }

\title{Personalized Filled-pause Generation with Group-wise Prediction Models}

\name{Yuta Matsunaga, Takaaki Saeki, Shinnosuke Takamichi, and Hiroshi Saruwatari\thanks{This work was supported by JST, Moonshot R\&D Grant Number JPMJPS2011.}}

\address{Graduate School of Information Science and Technology, The University of Tokyo, Japan.}

\abstract{In this paper, we propose a method to generate personalized filled pauses (FPs) with group-wise prediction models. Compared with fluent text generation, disfluent text generation has not been widely explored. To generate more human-like texts, we addressed disfluent text generation. The usage of disfluency, such as FPs, rephrases, and word fragments, differs from speaker to speaker, and thus, the generation of personalized FPs is required. However, it is difficult to predict them because of the sparsity of position and the frequency difference between more and less frequently used FPs. Moreover, it is sometimes difficult to adapt FP prediction models to each speaker because of the large variation of the tendency within each speaker. To address these issues, we propose a method to build group-dependent prediction models by grouping speakers on the basis of their tendency to use FPs. This method does not require a large amount of data and time to train each speaker model. We further introduce a loss function and a word embedding model suitable for FP prediction. Our experimental results demonstrate that group-dependent models can predict FPs with higher scores than a non-personalized one and the introduced loss function and word embedding model improve the prediction performance.
\\ \newline  }

\begin{document}

\maketitleabstract

\section{Introduction}
\label{sec:intro}

    Disfluency generation aims to generate human-like disfluent texts~\cite{Qader18disfluencyinsertion,yang20disfluencygeneration}. Compared with fluent text generation~\cite{brown20gpt3}, disfluent text generation has not been widely explored. Disfluency includes filled pauses (FPs), rephrases, and word fragments~\cite{schriberg94preliminariesTA}, and it is known that the tendency to use them varies from speaker to speaker~\cite{shriberg96disfluencies,schriberg94preliminariesTA,watanabe19japanesefiller}. Disfluency generation reproducing such individuality makes it possible to generate personalized disfluent texts and can be applied to spontaneous speech synthesis, which generates more human-like spontaneous speech than a reading-style one. In this research, we focus on spontaneous speech synthesis and address the disfluency generation reproducing individuality.
    
\begin{figure}[t]
        \centering
        \includegraphics[scale=0.32]{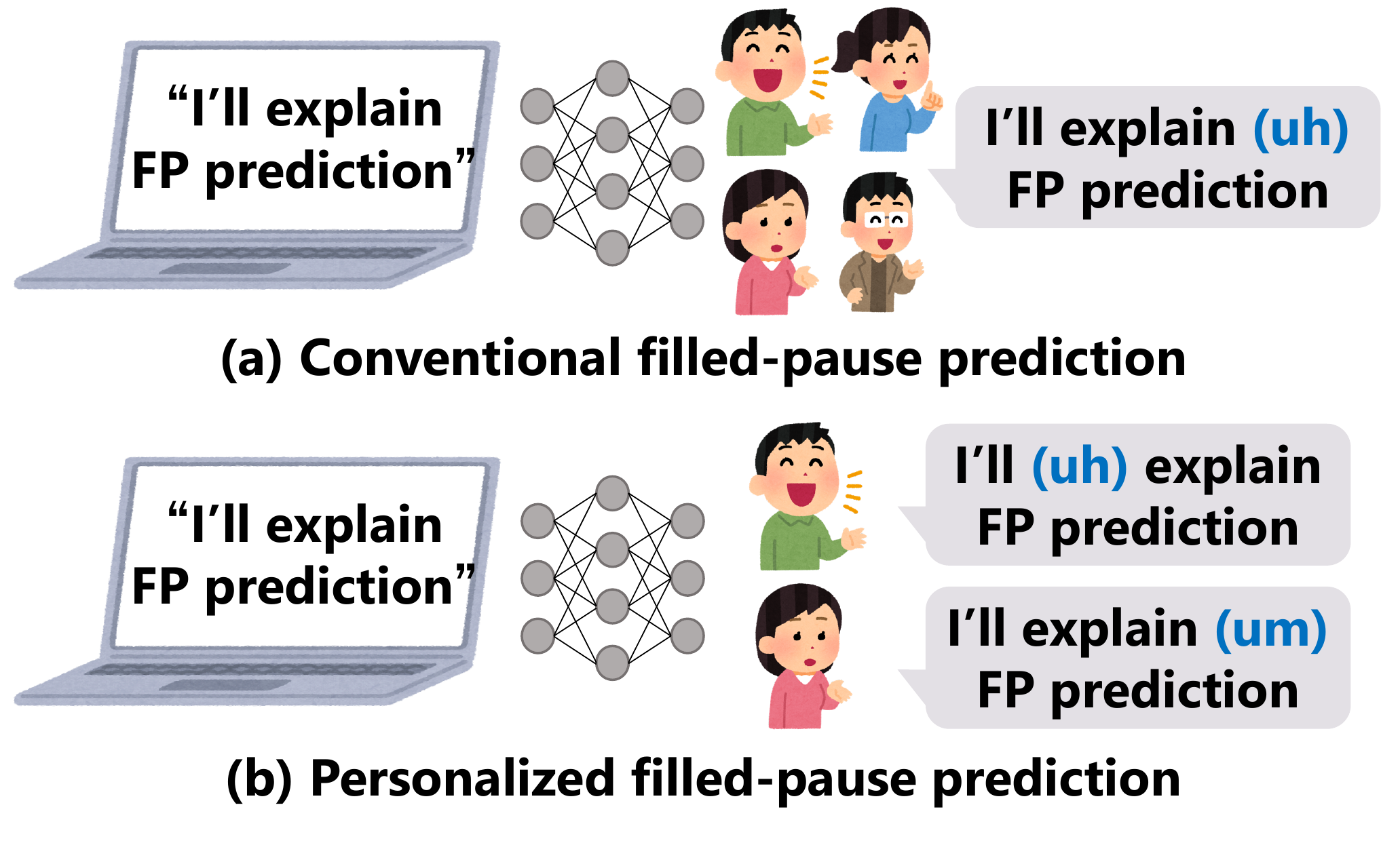}
        \vspace{-3mm}
        \caption{Personalized filled-pause (FP) prediction. In contrast to the aforementioned conventional method that predicts only FPs generalized among speakers, our proposed method further aims to predict FPs personalized to each speaker.}
        \vspace{-4mm}
        \label{fig:concept}
    \end{figure}

    FPs are defined as words that have a filling-in role~\cite{koiso01csjtranscription}, and there are various words for FPs~\cite{brown17listening,hirose06japanesefiller}. Such FPs are important because they have various effects on spontaneous speech. They play an important role in speech generation: planning~\cite{maclay59hesitation} and monitoring~\cite{levelt83monitoring}. They are also important to facilitate communication: speakers can indicate that they are searching for words~\cite{clark02usinguhum}, and listeners can understand the word quickly~\cite{fox99ohdiscourse}.
    The use of FPs influences the perception of the speaker's personality for listeners~\cite{gustafson21personalityinthemix}. 
    We, therefore, focus on FP generation to achieve FP-included spontaneous speech synthesis with these various effects.
    In addition, it is known that the position~\cite{shriberg96disfluencies} and words~\cite{schriberg94preliminariesTA,watanabe19japanesefiller} of FPs differ from speaker to speaker. Therefore, as shown in Figure~\ref{fig:concept}, we propose a personalized FP prediction method to reproduce not only whether each speaker uses FPs or not, but also how each speaker uses them.

    FP-included spontaneous speech synthesis consists of FP insertion and speech synthesis. The FP insertion model predicts or selects the position and word of FPs to generate disfluent texts from original fluent texts. The speech synthesis model generates the acoustic features of the fluent and FP parts from texts containing FPs. FP prediction is particularly difficult because of the sparsity of positions~\cite{ohta07languagemodelusingfillerprediction} and the bias of words (i.e. the frequency difference between more and less frequently used FP words). It is necessary to establish an FP prediction method to address these issues.

    In this paper, we propose a group-dependent FP prediction method using speaker grouping that can highly reproduce individuality. First, we divide speakers into groups on the basis of their tendency to use FPs. Then, we train group-dependent models by fine-tuning on the basis of the non-personalized model trained on multi-speaker data. This method does not require a large amount of spontaneous speech data for each speaker, nor the time to train each speaker-dependent model.
    
    The experimental results clarify that all the group-dependent models based on FP words have better F-scores: $0.456$ for positions and $0.288$ for words at most, compared with the non-personalized model with F-scores of $0.376$ for positions and $0.089$ for words. Moreover, almost all the models based on FP positions have better F-scores: $0.461$ for positions and $0.277$ for words at most than those of the non-personalized model.
    
    In addition, we introduce a loss function that takes into account sparsity and bias and a rich word embedding model, which improves F-scores for positions and words. The key contributions of this work are as follows:
    
    \begin{itemize}
        \item{We propose a method that constructs group-dependent models by grouping speakers on the basis of the tendency to use FPs and demonstrate that the performance of almost all the models is better than the non-personalized model. The group-dependent models and source implementation to train them are available at the github repository\protect\footnote{\url{https://github.com/ndkgit339/filledpause_prediction_group}}.}
        \item{We introduce a loss function suitable for FP prediction and a rich word embedding model and demonstrate that it improves the performance of the FP prediction models.}
    \end{itemize}

\section{Related work}
\label{sec:related_work}

    \subsection{FP-included speech generation} 
        Various studies have addressed FP-included speech generation from texts or fluent speech. \cite{yan2021adaspeech3,szekely19howtotrainfiller} have proposed methods to synthesize FP-included disfluent speech from FP-excluded fluent text. There is also a method to synthesize FP-included speech by using FPs' information as input~\cite{szekely19conversationalsynthfounddata}. In~\cite{adell08inserteditingterms}, the authors modeled the insertion of editing terms into fluent utterances and local prosodic changes. In addition, a number of studies have created an external module to predict FPs for FP-included speech synthesis~\cite{Wester15artifitialpersonality,gustafson21personalityinthemix,Cong21controllablecontextaware}. However, these studies have not attempted to improve the performance of the prediction or reproduce the FPs' individuality.
    
    \subsection{FP-included text prediction}
        A number of studies have focused on FP-included text prediction from FP-excluded fluent texts. \cite{Qader18disfluencyinsertion} proposed an algorithm using a probabilistic model to generate disfluent sentences from fluent sentences, but the prediction of FP words is simple. To construct the spoken language model, \cite{ohta07languagemodelusingfillerprediction} predicted the FPs' positions and words in order, but the scores of predicting each word are low. \cite{yamazaki20blstmfillerprediction} reported that simultaneously predicting positions and words improves performance. \cite{tomalin15latticebasedfillerprediction} also predicted FP word and position simultaneously using a lattice-based n-gram model. However, these methods cannot reproduce the diversity of FP words; the scores of the prediction of less frequent FPs are still low. In contrast to these studies, we propose a method to train prediction models reproducing the diversity of FP words.

\section{Method} \label{sec:method}
    To reproduce individuality in FP prediction, it is necessary to predict the positions and words of FPs more precisely for each speaker. First, we introduce a loss function that addresses the sparsity of positions and the bias of words and a rich word embedding model. Furthermore, we propose a method for building personalized prediction models.
            
    \subsection{Basic architecture and FP vocabulary} \label{ssec:method_basic}
        We construct a model that consists of two modules: a word embedding model and an embedding-to-FPtag model~\cite{yamazaki20blstmfillerprediction}. The word embedding model generates word embeddings for each morpheme from a sequence of morphemes segmented by morphological analysis of fluent text. We use a word embedding model that has been pre-trained on large-scale Japanese text data. The embedding-to-FPtag model predicts $14$ classes of no FP or $13$ FP words for each morpheme embedding. In this paper, following the previous work of~\cite{yamazaki20blstmfillerprediction}, we use a bidirectional long short-term memory (BLSTM)~\cite{graves05blstm} as an embedding-to-FPtag model and a cross entropy loss.
        
        We select the $13$~FP words by excluding any FP words used less frequently ($<$ $20$\%) among all speakers using the Corpus of Spontaneous Japanese (CSJ~\cite{maekawa04csj}) and cover approximately $83$\% of the FPs used by each speaker. Table~\ref{tab:fps} lists the FP words and example sentences.

\begin{table}[t]
\vspace{-1mm}
\footnotesize
\centering
\caption{List of FP words and example utterances\protect\footnotemark in CSJ (the lecture ID is A05M0058).}
\vspace{-0mm}
\label{tab:fps}
\begin{tabular}{l|l}
\hline
\multicolumn{1}{c|}{\begin{tabular}[c]{@{}c@{}}FP word\\Japanese (English)\end{tabular}} & \multicolumn{1}{c}{Example of an utterance}                   \\ \hline

\jp{えー} (ee)                & \begin{tabular}[c]{@{}l@{}}\jp{それをどう\textcolor{blue}{えー}判断するか}\\c.f.) how you \textcolor{blue}{uh} judge it\end{tabular} \\ \hline
\jp{え} (e)                 & \begin{tabular}[c]{@{}l@{}}\jp{\textcolor{blue}{え}大きく内容分けますと}\\c.f.) \textcolor{blue}{uh} roughly divide contents\end{tabular} \\ \hline
\multirow{2}{*}{\jp{ま} (ma)}                &  \multirow{4}{*}{\begin{tabular}[c]{@{}l@{}}\jp{\textcolor{blue}{あの}音声だとか言語\textcolor{blue}{ま}}\\\jp{そういう分野もございますし}\\c.f.) there are also areas such as\\\textcolor{blue}{um} speech, language, \textcolor{blue}{uh} and so on\end{tabular}} \\ 
 &  \\ \cline{1-1}
\multirow{2}{*}{\jp{あの} (ano)} & \\ 
 & \\ \hline
\multirow{2}{*}{\jp{あのー} (anoo)}              &  \multirow{4}{*}{\begin{tabular}[c]{@{}l@{}}\jp{\textcolor{blue}{まー}これ\textcolor{blue}{あのー}本当の意味}\\\jp{よく分からないんで}\\c.f.) I'm not sure what \textcolor{blue}{uh}\\this \textcolor{blue}{um} really means\end{tabular}} \\
 & \\ \cline{1-1}
\multirow{2}{*}{\jp{まー} (maa)}               &  \\
 & \\ \hline
\jp{えーと} (eeto)              & \begin{tabular}[c]{@{}l@{}}\jp{\textcolor{blue}{えーと}これはあのー部屋の}\\c.f.) \textcolor{blue}{um} this is uh the room\end{tabular} \\ \hline
\jp{あ} (a)                 & \begin{tabular}[c]{@{}l@{}}\jp{これが\textcolor{blue}{あ}副次的な効果として}\\c.f.) this is \textcolor{blue}{uh} a side effect of the\end{tabular} \\ \hline
\jp{あー} (aa)                & \begin{tabular}[c]{@{}l@{}}\jp{それから\textcolor{blue}{あー}建物の中にも}\\c.f.) then, in \textcolor{blue}{uh} the building\end{tabular} \\ \hline
\jp{ん} (n)                 & \begin{tabular}[c]{@{}l@{}}\jp{\textcolor{blue}{ん}何て言うんですかね}\\c.f.) \textcolor{blue}{uh} what can I say\end{tabular} \\ \hline
\jp{んー} (nn)                & \begin{tabular}[c]{@{}l@{}}\jp{\textcolor{blue}{んー}模型を現場に持ち込んで}\\c.f.) bring \textcolor{blue}{uh} a model to the site\end{tabular} \\ \hline
\jp{えっと} (etto)              & \begin{tabular}[c]{@{}l@{}}\jp{\textcolor{blue}{えっと}ただ小さい模型作って}\\c.f.) \textcolor{blue}{uh} just make a small model\end{tabular} \\ \hline
\jp{あーのー} (aanoo)             & \begin{tabular}[c]{@{}l@{}}\jp{\textcolor{blue}{あーのー}持っていかせないと}\\c.f.) \textcolor{blue}{uh} have to let him take it\end{tabular} \\ \hline
\end{tabular}
\vspace{-2mm}
\end{table}
\footnotetext{In the English translation of the example utterances, we set FP positions before the English words corresponding to the next words of the Japanese FPs and used ``uh'' and ``um''(if there were two or more FP words) as FP words.}

    \subsection{Weighted cross entropy loss} \label{ssec:method_weighted_loss}
        Since FPs tend to be sparse in positions and biased in words, there is a problem in that the model predicts only no FPs or highly frequent FPs~\cite{ohta07languagemodelusingfillerprediction}. Therefore, we use weighted cross entropy loss to build a model to predict even less frequent FPs precisely~\cite{yan2021adaspeech3}. The loss weights of the $14$ predicted classes are the reciprocals of their frequencies in the training data so that the losses of less frequent FPs have large values.
    
    \subsection{Rich word embedding model} \label{ssec:method_bert}
        A previous study~\cite{yamazaki20blstmfillerprediction} used fastText~\cite{bojanowski2017fasttext}, which is a lightweight model that generates word representations, as the word embedding model. Whereas fastText calculates the word embedding without considering context information (e.g., position and neighboring words), Bidirectional Encoder Representations from Transformers (BERT)~\cite{devlin19bert} calculates the embeddings from the entire input texts considering context information. As previous studies have shown the relationship between subsequent clause length and FP usage~\cite{watanabe19japanesefiller}, there is a long context dependency for FP insertion. Since BERT is more appropriate than fastText to capture this, we compare their prediction performances.
        
    \subsection{Personalized FP prediction model} \label{ssec:method_group}
\begin{figure}[t]
            \centering
            \includegraphics[scale=0.35]{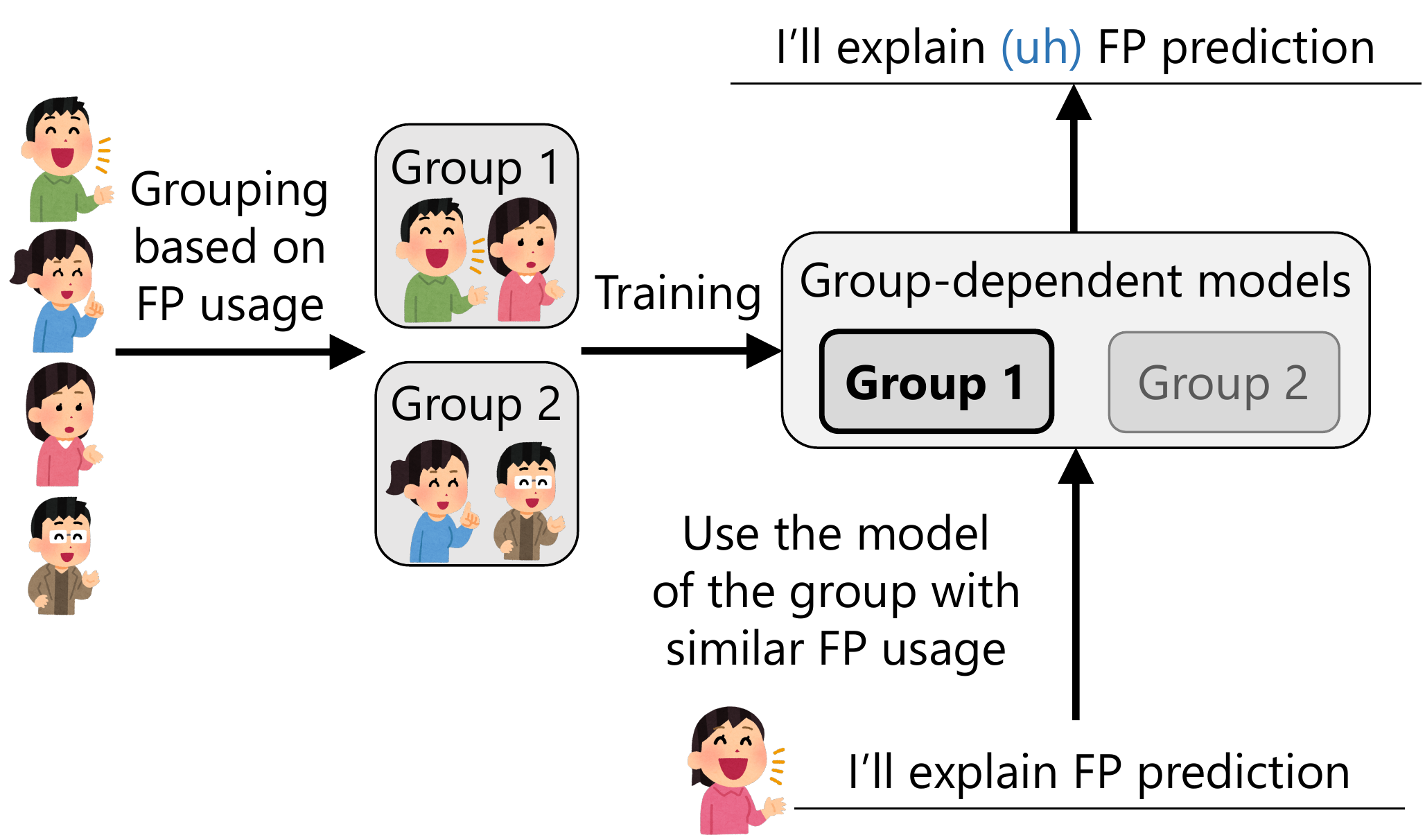}
            \vspace{-3mm}
            \caption{Constructing group-dependent models on the basis of tendency of FP usage}
            \vspace{-4mm}
            \label{fig:method}
        \end{figure}

        This study aims to develop an FP prediction method taking into account individuality. We can construct speaker-dependent models using a sufficient amount of spontaneous speech data of target speakers; however, it requires a large amount of time and data to train each target speaker's model. To address this issue, as shown in Figure~\ref{fig:method}, we propose a group-dependent model training method based on speaker grouping as an efficient way to train models that reproduce the individuality of FPs. First, we train a non-personalized model using a multi-speaker spontaneous speech corpus that contains transcriptions and FP annotations. Then, we train group-dependent models by fine-tuning the non-personalized model. We also train speaker-dependent models for comparison with these models.
        
        \textbf{Group-dependent model.} 
        We use a grouping method like hierarchical clustering to group speakers and perform fine-tuning to update the parameters of the sequence-to-sequence model with the data of each group using the parameters of the non-personalized model as initial values. We use the model of the group that has an FP tendency closest to the target speaker for the inference. This method does not train the prediction model of each target speaker. Therefore, we can reduce the cost to collect a large amount of spontaneous speech data of each speaker. Moreover, we do not require time to train the model of each speaker.
        
        \textbf{Speaker-dependent model.} 
        To compare with the group-dependent model, we adapt the non-personalized model to the speakers. We use mid-size ($3.5$--$5.0$ hours) spontaneous speech data. We manually search lecture videos on the web and transcribe and annotate texts in accordance with the rules of the CSJ. The data includes transcribed fluent texts as well as FP words, FP tags, and phrase timings. We train speaker-dependent models with this data by fine-tuning the non-personalized model in the same way as the group-dependent models.

\section{Experiment} \label{sec:exp}

    \subsection{Experimental setting} \label{ssec:exp_setting}
        
        \subsubsection{Evaluation criteria}
            We first describe the criteria of the evaluations. To score the model to predict FPs' positions and words, we used precision, recall, F-score, and specificity. First, for positions, we defined precision as the rate of positions actually having FPs out of those predicted to have FPs and recall as the rate of positions predicted to have FPs out of those actually having FPs. Then, we calculated the F-scores from these scores. The specificity was the rate of positions predicted to have no FPs out of the positions actually having no FPs. For each FP's word, we calculated precision, recall, F-score, and specificity in the same way. To score the prediction for FP words, we used the average of each word's score weighted by the frequency of each FP.
        
        \subsubsection{Training} \label{sssec:exp_setting_train}
            In the following experiments, we used the data of $137$~speakers from the CSJ. As the utterance unit, we used the breath groups separated by silence for $0.2$~seconds or longer. We separated each breath group into morphemes using Sudachi~\cite{takaoka18sudachi} for fastText and Juman++~\cite{morita15jumanpp} for BERT. For the word embedding model, we used fastText with the dimension set to $300$\footnote{\url{https://fasttext.cc/docs/en/crawl-vectors.html}} and BERT with the version of the LARGE WWM published by the Kurohashi-Chu-Murawaki labs at Kyoto University\footnote{\url{https://nlp.ist.i.kyoto-u.ac.jp/?ku_bert_japanese}}. These models were pre-trained on large Japanese text data on Common Crawl and Wikipedia for fastText and on Wikipedia for BERT. We used BLSTM as the embedding-to-FPtag model and set the number of hidden layers and hidden size to $1$ and $1024$, respectively. We apply gradient clipping with the maximum of the norm set to $0.5$ and set the batch size to $32$. Unless otherwise stated, we set the learning rate to 1.0×\( 10^{-5} \) and trained the models for $60000$~steps.

        \subsubsection{Cross validation} \label{sssec:exp_setting_crossval}
            To evaluate the results of the prediction by the models, we applied cross validation. We divided the speakers into $10$~sets. Then, we trained the models using $9$ of these sets as training data with the remaining set for evaluation. The average of the evaluation scores obtained by repeating this process $10$~times was the evaluation score of the model. We considered the missed scores as $0.0$ and calculated the average score.

        \subsubsection{Method of grouping speakers}
            We describe the method of grouping speakers. First, for clustering by FP word usage, we calculated the rate of usage of each FP word by dividing the number of each FP word usage by the total number of FP usage for each speaker. Then, we applied hierarchical clustering using Euclidean distance and Ward's method~\cite{ward63clustering}. We compared the results of the clustering by using a number of distance thresholds and then set the threshold to $1.0$ which seems to have the largest difference between clusters. We then classified the speakers into $4$~classes. Next, for clustering by positional tendency, we use 1) head of the sentence, 2) intra-sentence and boundary of breath group, 3) intra-sentence and middle of breath group, and 4) end of the sentence as FP positions, and calculated the rate of each FP position usage by dividing the number of each FP position usage by the total number of FPs for each speaker. Then, we applied hierarchical clustering using Euclidean distance and Ward's method. On the basis of the same criterion as the clustering by FP word, we set the distance threshold to $1.7$ and then classified the speakers into four classes on the basis of the distance.

    \subsection{Weighted cross entropy loss}
    \label{ssec:exp_weighted_loss}
        To investigate whether the weighted cross entropy loss is effective for predicting FPs, we first compared the prediction scores between the equal and weighted losses. In this evaluation, for a word embedding model, we used fastText which is the conventional model in~\cite{yamazaki20blstmfillerprediction}. The appropriate hyper-parameter settings with and without the weighted loss function were different because the objective loss functions of the training were different. In this evaluation, as hyper-parameters suitable for both settings, we set the learning rate to 1.0×\( 10^{-3} \), multiplied $0.1$ every $100000$~steps, and trained the model for $200000$~steps.
        
        Table~\ref{tab:weighted_loss} lists the results. We can see that the F-scores are higher for both positions and words in weighted than in equal, indicating that using loss weights improved the performance of the prediction of positions and words. The precision and recall of the model with weights are lower and higher, respectively, for both positions and words. This suggests that introducing the weights makes the model actively insert FPs and improves the recalls, but it increases the number of mistakes. This is consistent with the result that the position's specificity is decreasing.
            
        On the basis of this result, we used the weighted cross entropy loss in the following experiments.
        
        \begin{table}[t]
\vspace{-3mm}
\footnotesize
\centering
\caption{Evaluation of weighted cross entropy loss}
\label{tab:weighted_loss}
\begin{tabular}{ll|l|l}
\hline
\multicolumn{2}{l|}{Criterion}                            & Equal & Weighted \\ \hline
\multicolumn{1}{c}{\multirow{4}{*}{Position}} & Precision & \textbf{0.307}    & 0.292 \\
\multicolumn{1}{c}{}                          & Recall    & 0.094             & \textbf{0.287} \\
\multicolumn{1}{c}{}                          & F-score   & 0.143             & \textbf{0.288} \\
\multicolumn{1}{c}{}                          & Specificity & \textbf{0.997}    & 0.989 \\ \hline
\multirow{4}{*}{Word}                         & Precision & \textbf{0.088}    & 0.078          \\
                                              & Recall    & 0.028             & \textbf{0.047} \\
                                              & F-score   & 0.042             & \textbf{0.054} \\
                                              & Specificity & 0.999             & 0.999 \\ \hline
\end{tabular}
\end{table}

    \subsection{Rich word embedding model} \label{ssec:exp_bert}
        To investigate whether the BERT, a rich word embedding model is effective for predicting FPs, we compared the prediction scores of fastText and BERT when used as the word embedding model. In this evaluation, we used the hyper-parameters described in Section~\ref{sssec:exp_setting_train}.

        Table~\ref{tab:fasttext_bert} lists the results. We can see that BERT has higher F-scores than fastText, which indicates that the prediction performance is improved by using BERT as a word embedding model.
        
        On the basis of this result, BERT was used as the word embedding model in the following experiments.
        
        \begin{table}[t]
\vspace{-3mm}
\footnotesize
\centering
\caption{Comparison of fastText (lightweight embedding model) and BERT (rich embedding model)}
\label{tab:fasttext_bert}
\begin{tabular}{ll|l|l}
\hline
\multicolumn{2}{l|}{Criterion}                            & fastText        & BERT \\ \hline
\multicolumn{1}{c}{\multirow{4}{*}{Position}} & Precision & 0.237           & \textbf{0.254} \\
\multicolumn{1}{c}{}                          & Recall    & \textbf{0.756}  & 0.732 \\
\multicolumn{1}{c}{}                          & F-score   & 0.360           & \textbf{0.376} \\
\multicolumn{1}{c}{}                          & Specificity & 0.961           & \textbf{0.964} \\ \hline
\multirow{4}{*}{Word}                         & Precision & 0.069           & \textbf{0.070} \\
                                              & Recall    & 0.138           & \textbf{0.149} \\
                                              & F-score   & 0.065           & \textbf{0.089} \\
                                              & Specificity & \textbf{0.996}           & 0.994 \\ \hline
\end{tabular}
\end{table}

    \subsection{Comparison of speaker-close and speaker-open prediction} \label{ssec:exp_open_close}
        We compared the prediction scores for speakers included (i.e. speaker-close) and not included (i.e. speaker-open) in the training data. In this evaluation, we used BERT as a word embedding model and the weighted cross entropy loss, which were proven to perform better in previous evaluations. In the cross validation of this evaluation, we also split the $9$~sets for training, described in Section~\ref{sssec:exp_setting_crossval}, to training and validation data in a ratio of approximately $9$:$1$ with the speaker-close condition. We used that validation data for speaker-close evaluation and the $1$~remaining set for speaker-open evaluation.

        Table~\ref{tab:open_close} lists the results. We can see that the speaker-close prediction has higher F-scores than speaker-open, indicating that the speaker-close prediction has better performance for both position and word. However, the difference between F-scores is only about $0.004$ for positions and $0.005$ for words, indicating that the speaker-open prediction achieves comparable performance to the speaker-close one. Therefore, we can use the prediction models to predict the FPs of the unseen speaker. Moreover, we can use the non-personalized model to predict FPs of the unseen speaker when it is not important to predict the personalized FPs but the non-personalized FPs are required. The value of specificity is close to $1.0$, indicating that FPs are rarely inserted in positions where there are actually no FPs. This is also true for the other results.
        
        \begin{table}[t]
\vspace{-3mm}
\footnotesize
\centering
\caption{Comparison of speaker-open and speaker-close evaluation}
\label{tab:open_close}
\begin{tabular}{ll|l|l}
\hline
\multicolumn{2}{l|}{Criterion}                            & Close          & Open           \\ \hline
\multicolumn{1}{c}{\multirow{4}{*}{Position}} & Precision & \textbf{0.264}          & 0.263 \\
\multicolumn{1}{c}{}                          & Recall    & \textbf{0.728} & 0.714          \\
\multicolumn{1}{c}{}                          & F-score   & \textbf{0.387} & 0.383          \\
\multicolumn{1}{c}{}                          & Specificity & 0.966 & 0.966          \\ \hline
\multirow{4}{*}{Word}                         & Precision & \textbf{0.073} & 0.071          \\
                                              & Recall    & \textbf{0.162} & 0.147          \\
                                              & F-score   & \textbf{0.096} & 0.091          \\
                                              & Specificity & 0.994 & 0.994          \\ \hline
\end{tabular}
\end{table}
        
    \subsection{Personalized FP prediction model} \label{ssec:exp_group}
        We first show the results of hierarchical clustering. Then, we describe the prediction scores of the two types of personalized models: group-dependent ones and speaker-dependent ones. Moreover, we present the scores of the prediction of each FP and the distribution of the scores across speakers. Finally, we describe the results of the prediction on lecture data of $2$~speakers.
        We trained the group-dependent models and speaker-dependent models for $10000$~steps. In the cross validation of group-dependent models, we set the number of speaker partitions to $5$ unlike the other experiments, since the amount of data in each group was smaller than before. For speaker-dependent models, as described in Section~\ref{ssec:method_group}, we used the lecture data of the University of Tokyo available on YouTube\footnote{\url{https://youtube.com/playlist?list=PLHxBhbJJasnfX6oBrkTygP8we61wEcRha}} for $2$~speakers. The test data for each speaker was $20$~paragraphs. We split the rest of the data into training and validation data in a ratio of approximately $9$:$1$.

        \subsubsection{Result of clustering} \label{sssec:exp_result_clustering}
            We describe the characteristics of the classes into which the speakers were classified by clustering on the basis of their tendency to use FPs.

            \begin{figure}[t]
\centering
\includegraphics[scale=0.2, trim=5.0 20.0 5.0 0.0]{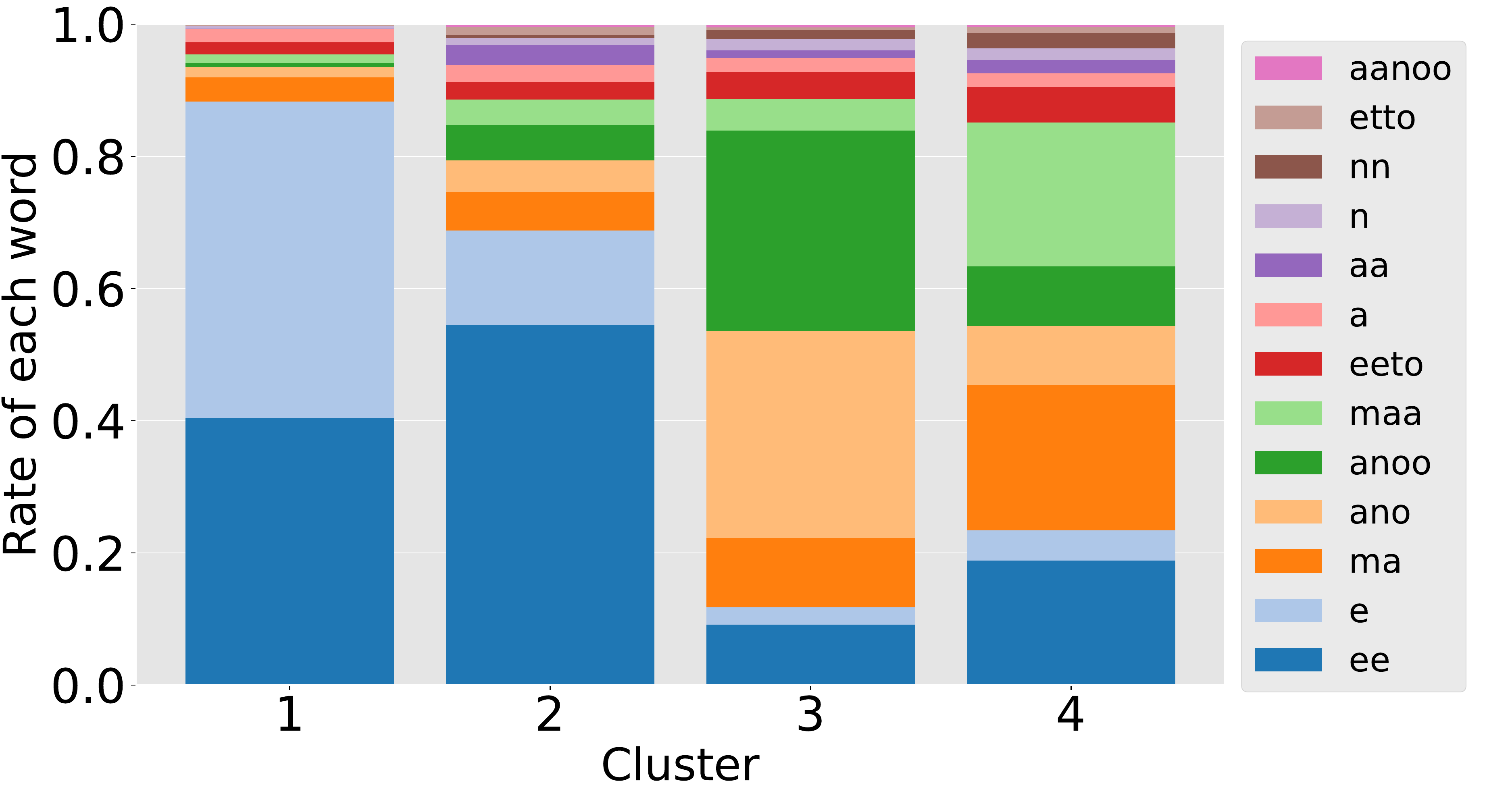}
\vspace{-2mm}
\caption{Rate of each FP word in each cluster by FP word usage}
\label{fig:clustering_word}
\vspace{-3mm}
\end{figure}

\begin{figure}[t]
\centering
\includegraphics[scale=0.26, trim=5.0 20.0 5.0 0.0]{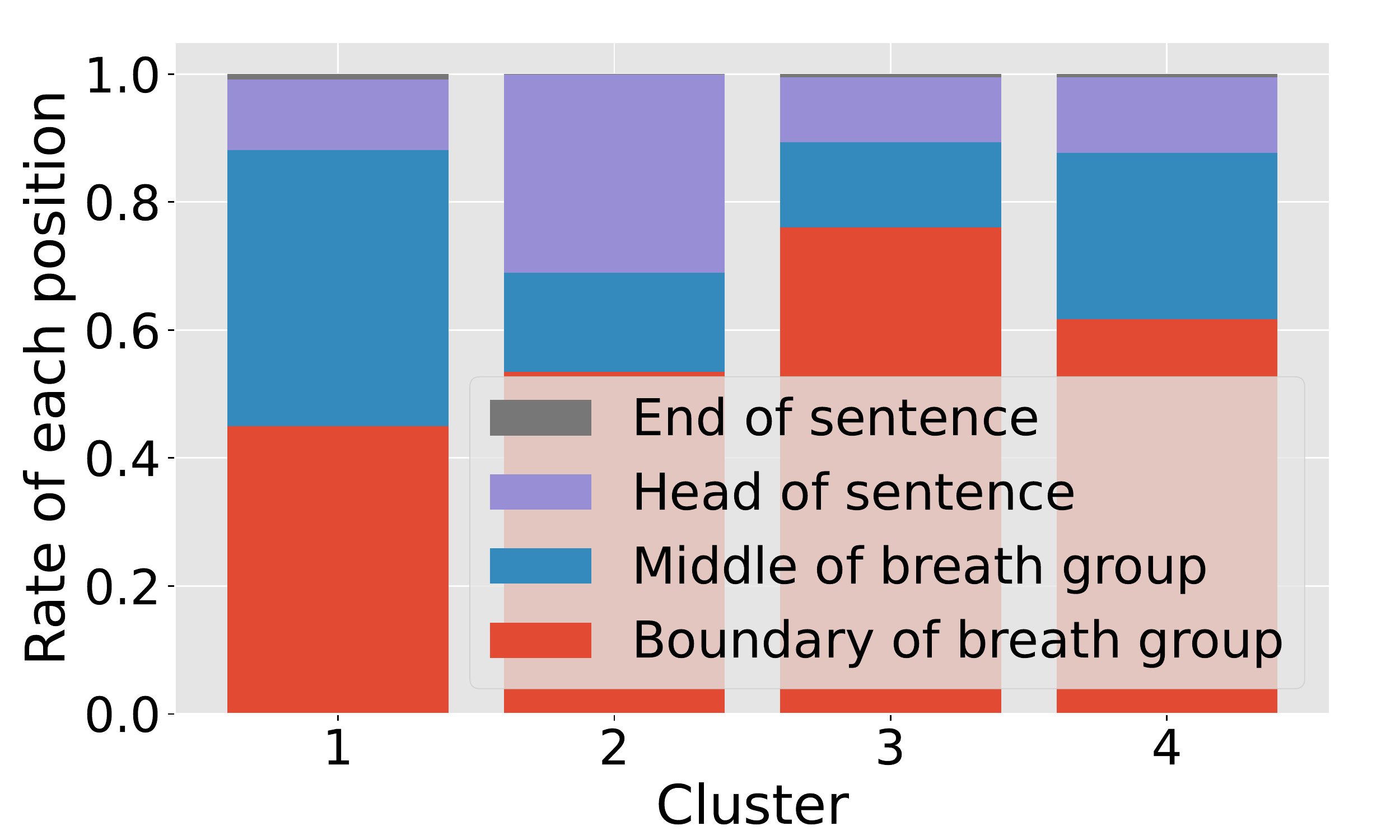}
\vspace{-2mm}
\caption{Rate of each FP position in each cluster by FP position usage}
\label{fig:clustering_position}
\vspace{-3mm}
\end{figure}

            Figure~\ref{fig:clustering_word} shows the results of calculating each FPs' word rate for each class by FP words. Cluster~$1$, $2$, $3$, and $4$ contain $18$, $55$, $25$, and $39$~speakers respectively. Frequently used FPs are ``ee'' in Cluster~$2$, ``ano'' and ``anoo'' in Cluster~$3$, ``ma'' and ``maa' in Cluster~$4$, whereas "ee" is frequently used in Cluster~$1$ and FPs other than "ee" and "e" are rarely used.
            Figure~\ref{fig:clustering_position} shows the results of calculating each FPs' position rate for each class by FP positions. Clusters~$1$, $2$, $3$, and $4$ contain $50$, $13$, $27$, $47$~speakers respectively. FPs are used more frequently in the middle of the breath group in Cluster~$1$, at the head of sentences in Cluster~$2$, and at the beginning of the breath group in Cluster~$3$, whereas Cluster~$4$ shows an average tendency among all the classes.
            
        \subsubsection{Evaluation of group-dependent models} \label{sssec:exp_result_group}
            To investigate the effectiveness of prediction by the proposed group-dependent models, we compared the prediction scores of these models with those of the non-personalized model.
            
            Table~\ref{tab:class_word} lists the results. We can see that the F-scores of all the group-dependent models based on FP words are higher than that of the non-personalized model for both positions and words, indicating that the grouping of speakers by FP words proposed in this paper improves the performance of the prediction. Table~\ref{tab:class_position} shows that the F-scores of the group-dependent models are higher than that of the non-personalized model, except for the position's F-score of Cluster~$2$, indicating that the grouping of speakers by FP positions proposed in this paper improves the performance for almost all the group-dependent models.
            
\begin{table}[t]
\vspace{-3mm}
\footnotesize
\centering
\caption{Evaluation of group-dependent models based on FP words (NP represents the non-personalized model.)}
\label{tab:class_word}
\begin{tabular}{l|l|llll}
\hline
\multirow{2}{*}{\begin{tabular}[c]{@{}l@{}}Criterion\\ (F-score)\end{tabular}}    & \multirow{2}{*}{NP} & \multicolumn{4}{c}{Group-dependent (word)}      \\ \cline{3-6} 
                              &                          & 1     & 2     & 3     & 4     \\ \hline
\multicolumn{1}{c|}{Position} & 0.376                    & \textbf{0.454} & \textbf{0.456} & \textbf{0.427} & \textbf{0.390} \\
Word                          & 0.089                    & \textbf{0.284} & \textbf{0.288} & \textbf{0.248} & \textbf{0.196} \\ \hline
\end{tabular}
\end{table}

\begin{table}[t]
\vspace{-3mm}
\footnotesize
\centering
\caption{Evaluation of group-dependent models based on FP positions (NP represents the non-personalized model.)}
\label{tab:class_position}
\begin{tabular}{l|l|llll}
\hline
\multirow{2}{*}{\begin{tabular}[c]{@{}l@{}}Criterion\\ (F-score)\end{tabular}}    & \multirow{2}{*}{NP} & \multicolumn{4}{c}{Group-dependent (position)}  \\ \cline{3-6} 
                              &                          & 1     & 2     & 3     & 4     \\ \hline
\multicolumn{1}{c|}{Position} & 0.376                    & \textbf{0.461} & 0.323 & \textbf{0.413} & \textbf{0.444} \\
Word                          & 0.089                    & \textbf{0.277} & \textbf{0.212} & \textbf{0.158} & \textbf{0.237} \\ \hline
\end{tabular}
\end{table}

        \subsubsection{Evaluation of prediction for each FP} \label{sssec:exp_result_eachfp}
            To investigate whether the group-dependent models can reproduce the diversity of FP words, we show the prediction scores of each FP.
            
\begin{figure}[t]
\centering
\includegraphics[scale=0.36, trim=5.0 30.0 0.0 0.0]{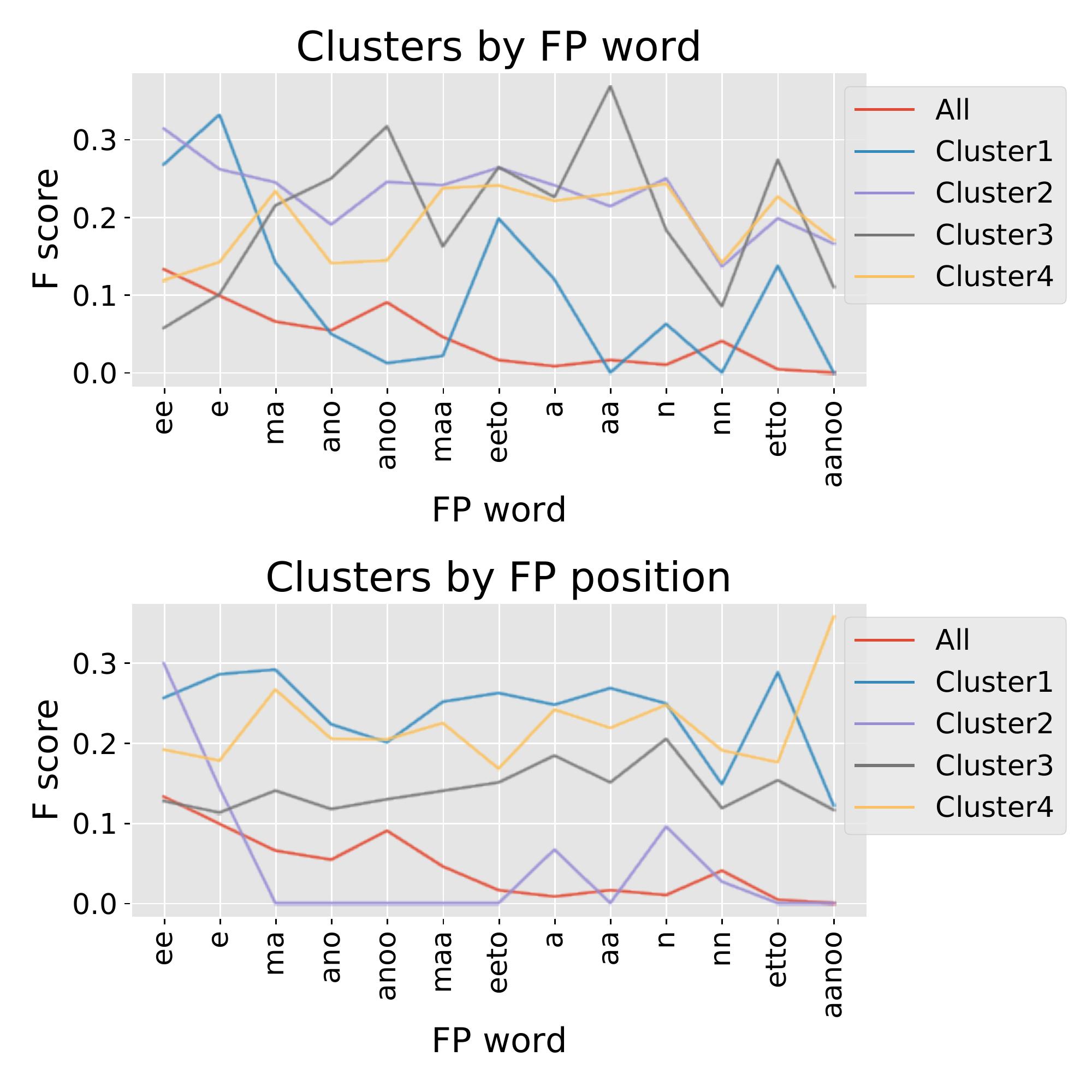}
\caption{F-score for each FP word on group-dependent model}
\label{fig:each_FP}
\vspace{-3mm}
\end{figure}

            Figure~\ref{fig:each_FP} shows the F-scores of the prediction on each FP word in the group-dependent models by FP words and that in the group-dependent models by FP positions. The horizontal axis from left to right represents more to less frequent FP words, respectively, in the corpus. The word's Cluster~$2$ and position's Clusters~$1$ and $4$ models have higher F-scores than the non-personalized model for all the FP words. In contrast to the non-personalized model, which predicts highly and less frequent FPs more and less precisely, respectively, the aforementioned models have F-scores for less frequent FPs close to that of frequent FPs, indicating that these models can reproduce the diversity of FP words. The word's Clusters~$3$ and $4$, and position's Cluster~$3$ models have better scores than the non-personalized model for FPs other than "ee," and the scores for less frequent FPs are high, also indicating the ability to reproduce the diversity. The model for word's Cluster~$1$ model has lower scores than the non-personalized model for a number of FPs. In addition, the position's Cluster~$2$ model has low F-scores of 0 for $8$~FPs.
            
        \subsubsection{Evaluation of predictions for each speaker} \label{sssec:exp_result_eachspk}
            To investigate the prediction performance of the group-dependent models for each speaker, we show the distribution of the prediction scores across speakers.
            
\begin{figure}[t]
\centering
\includegraphics[scale=0.36, trim=5.0 30.0 0.0 0.0]{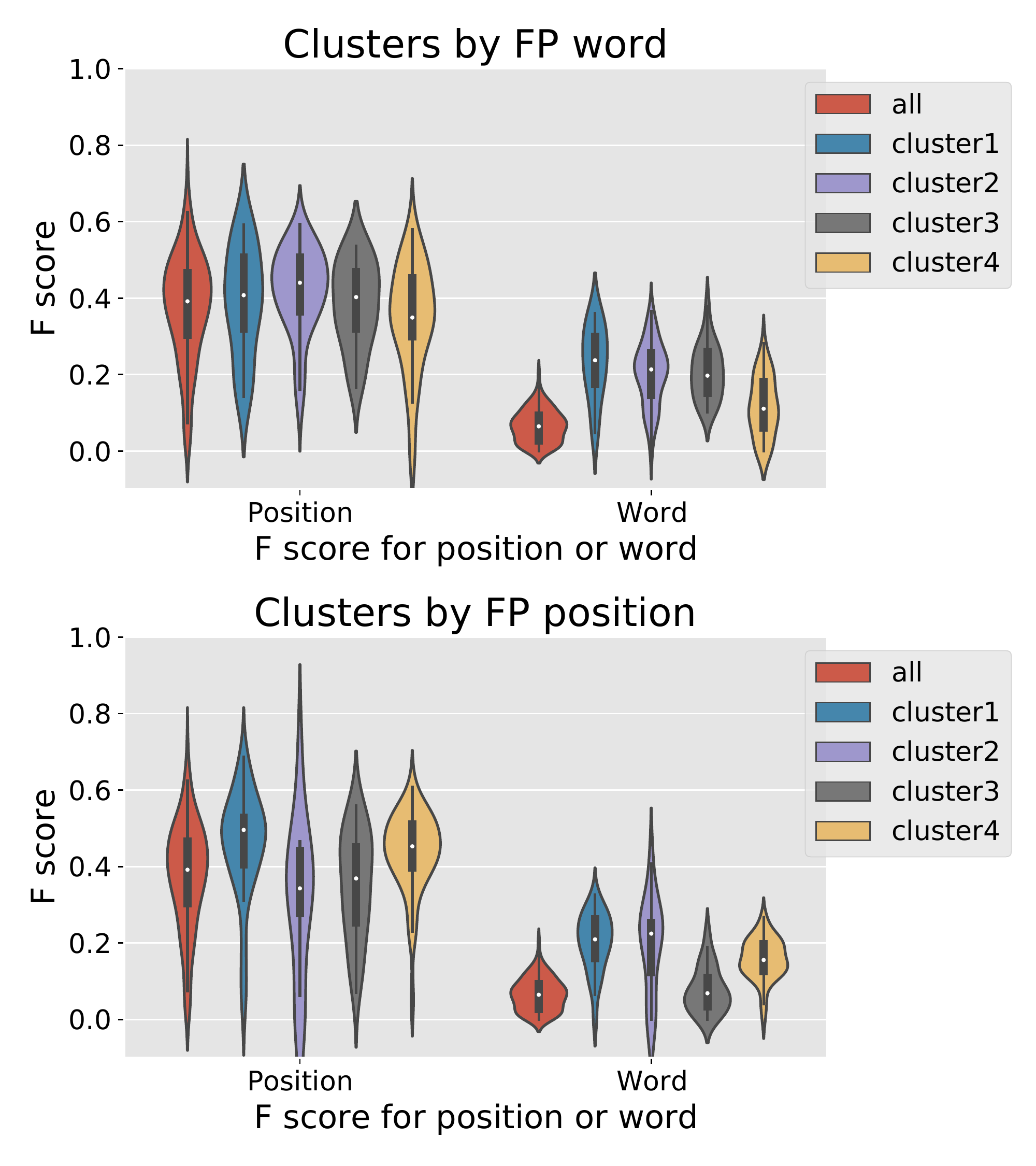}
\caption{F-score for each speaker on group-dependent model}
\label{fig:each_spk}
\vspace{-3mm}
\end{figure}

            We describe the distribution of each speaker's prediction score by the group-dependent models. The cross validation was not performed, but the data was divided into training, validation, and test data in an approximate ratio of $3$:$1$:$1$ under the speaker-close condition, and the results were evaluated using the test data. Figure~\ref{fig:each_spk} shows the results. The F-scores for each cluster of positions and words are widely distributed, indicating that grouping speakers improves the performance on average, but the tendency of improvement differs from speaker to speaker. Therefore, further research is needed to investigate which speakers have a worse prediction performance and construct prediction models that perform well for them.
            
        \subsubsection{Evaluation on lecture data} \label{sssec:exp_result_lec}
            To investigate whether the models can be adapted to speakers, we evaluated the prediction scores of the speaker-dependent models. We also compared the prediction performance of the non-personalized, group-dependent, and speaker-dependent models on the lecture data of $2$~speakers.
            
            Tables~\ref{tab:speaker_a} and \ref{tab:speaker_b} list the results of speakers~A and B, respectively. We can see that speaker-dependent models have lower scores than the non-personalized model for both speakers, indicating that it is difficult to adapt the prediction model to speakers. A possible reason is that the variation of the usage tendency within each speaker is large and the tendency differs between the training and evaluation data.
            
            In Table~\ref{tab:speaker_a}, we can see that the score of the non-personalized model is the highest, followed by the speaker-dependent model, and the group-dependent models have the lowest score. In Table~\ref{tab:speaker_b}, we can see that the score of the non-personalized model is the highest, followed by the group-dependent models, and the speaker-dependent model has the lowest score. For speakers in which the non-personalized model has the best performance and a high prediction score was same for both speaker~A and B, we can use the non-personalized model for inference.

        \subsubsection{Discussion} \label{sssec:exp_discuss}
            The experimental results demonstrated that grouping speakers on the basis of their FP usage enabled the construction of group-dependent models with higher prediction F-scores than the non-personalized model. One possible reason for this is that grouping by FP usage reduces the variation in usage tendency within the data group, which makes the model training easier. Another possible reason is that the difference in tendency between the training and evaluation data is reduced. This indicates the effectiveness of our experimental results because the prediction model of the group close to the target speaker's data is actually used to predict unseen speakers.
            
            In the evaluation using the lecture data, the F-score of the non-personalized model was the highest. Considering the results in Figure~\ref{fig:each_spk}, which shows that the scores of the prediction models differed among speakers, suggesting that the $2$~speakers in this experiment have particularly worse prediction performance by the group-dependent models. Considering that speaker~A has low scores for all the models, one possible reason is that speaker~A has a particularly unusual usage tendency. Considering that speaker~B has the best score for the non-personalized model, one possible reason is that speaker~B has a usage tendency close to the common tendency among all the speakers. For such speakers, we can use the non-personalized model for inference.


\begin{table}[t]
\vspace{-3mm}
\footnotesize
\centering
\caption{Evaluation on lecture data of speaker~A (NP represents the non-personalized model.)}
\label{tab:speaker_a}
\begin{tabular}{l|l|l|l|l}
\hline
\begin{tabular}[c]{@{}l@{}}Criterion\\ (F-score)\end{tabular} & NP & \begin{tabular}[c]{@{}l@{}}Word\\ (Cluster 4)\end{tabular} & \begin{tabular}[c]{@{}l@{}}Position\\ (Cluster 2)\end{tabular} & \multicolumn{1}{c}{Speaker} \\ \hline
\multicolumn{1}{c|}{Position}                                 & 0.243   & 0.137                                                     & 0.114                                                         & 0.146                       \\
Word                                                          & 0.061   & 0.016                                                     & 0.018                                                         & 0.057                       \\ \hline
\end{tabular}
\end{table}

\begin{table}[t]
\vspace{-3mm}
\footnotesize
\centering
\caption{Evaluation on lecture data of speaker~B (NP represents the non-personalized model.)}
\label{tab:speaker_b}
\begin{tabular}{l|l|l|l|l}
\hline
\begin{tabular}[c]{@{}l@{}}Criterion\\ (F-score)\end{tabular} & NP & \begin{tabular}[c]{@{}l@{}}Word\\ (Cluster 4)\end{tabular} & \begin{tabular}[c]{@{}l@{}}Position\\ (Cluster 1)\end{tabular} & \multicolumn{1}{c}{Speaker} \\ \hline
\multicolumn{1}{c|}{Position}                                 & 0.384   & 0.302                                                     & 0.366                                                         & 0.212                       \\
Word                                                          & 0.158   & 0.070                                                     & 0.117                                                         & 0.027                       \\ \hline
\end{tabular}
\end{table}

\section{Conclusion} \label{sec:conclusion}
    To achieve FP prediction to reproduce individuality, we proposed a method to construct group-dependent models with higher scores than the general model by grouping speakers on the basis of their FP usage. This method made it possible to predict the target speaker's FP without learning the speaker-dependent model of the target speaker each time. Furthermore, we introduced a weighted loss function to address the sparsity of FP positions and the bias of FP words and a rich word embedding model, and demonstrated that the performance of the prediction was improved. However, since we found that the prediction performance varied among speakers, we need to investigate which speakers have worse prediction performance and address performance improvement for those speakers. Moreover, our future work will involve synthesizing spontaneous speech containing FPs predicted by the group-dependent models proposed in this paper and subjectively evaluating individuality.

\section{Bibliographical References}\label{reference}

\bibliographystyle{lrec2022-bib}
\bibliography{lrec2022-example}


\end{document}